\documentstyle[prc,aps,epsfig,multicol]{revtex}
\draft

\newcommand{\ub}{\bar u}
\newcommand{\db}{\bar d}
\newcommand{\qb}{\bar q}
\newcommand{\sbar}{\bar s}
\newcommand{\pt}{p_{\rm T}}
\newcommand{\vpp}{\vec{\pt}}

\begin{document}
\title{Flavor ordering of elliptic flows at high transverse momentum}
\author{Zi-wei Lin and C.M. Ko}
\address{Cyclotron Institute and Physics Department,
Texas A\&M University, College Station, Texas 77843-3366}
\maketitle
\begin{abstract}
Based on the quark coalescence model for the parton-to-hadron phase
transition in ultra-relativistic heavy ion collisions, we relate the 
elliptic flow ($v_2$) of high $\pt$ hadrons to that of high $\pt$ quarks.  
For high $\pt$ hadrons produced from an isospin symmetric and 
quark-antiquark symmetric partonic matter, magnitudes of their
elliptic flows follow a flavor ordering as 
$(v_{2,\pi}=v_{2,N}) > (v_{2,\Lambda}=v_{2,\Sigma}) 
> v_{2,K} > v_{2,\Xi} > (v_{2,\phi}=v_{2,\Omega})$ 
if strange quarks have a smaller elliptic flow than light quarks. 
The elliptic flows of high $\pt$ hadrons further follow a simple quark 
counting rule if strange quarks 
and light quarks have same high $\pt$ spectrum and coalescence probability.
\end{abstract}
\vspace{0.3cm}
\vspace{0.3cm}

\begin{multicols}{2}

Elliptic flow in heavy ion collisions is a measure of the azimuthal asymmetry 
of particle momentum distributions in the plane perpendicular to the 
beam direction. It results from the initial spatial asymmetry in the 
transverse plane in non-central collisions and is thus sensitive to 
the properties of the dense matter formed during the initial stage 
of heavy ion collisions
\cite{Ollitrault:1992bk,Sorge:1996pc,Rqmd,Danielewicz:1998vz,Teaney:v2,Kolb:2000fh,Huovinen:2001cy,Kolb:2001qz}.  
There have been extensive experimental
\cite{Barrette:1994xr,Appelshauser:1998dg,Ackermann:2001tr,Lacey:2001va,Back:2002gz,Adler:2002pu} and theoretical 
\cite{Ollitrault:1992bk,Sorge:1996pc,Rqmd,Danielewicz:1998vz,Zheng:1999gt,Teaney:v2,Kolb:2000fh,Huovinen:2001cy,Kolb:2001qz} 
studies of elliptic flow in heavy ion collisions at various energies.
For heavy ion collisions at the Relativistic Heavy Ion Collider (RHIC), 
the elliptic flow has been measured as functions of the centrality of 
collisions \cite{Ackermann:2001tr,Lacey:2001va,Back:2002gz,Adler:2002pu}, 
as well as the particle transverse momentum 
\cite{Ackermann:2001tr,Lacey:2001va,Adler:2002pu} 
and pseudo-rapidity \cite{Back:2002gz}. 
Theoretical studies indicate that these experimental results provide
not only information on the equation of state of nuclear matter at
high density and temperature 
\cite{Teaney:v2,Kolb:2000fh,Huovinen:2001cy,Kolb:2001qz} 
but also on the scattering cross section of partons produced in the collisions
\cite{Zhang:1999rs,Molnar:v2,Zabrodin:2001rz,Lin:2001zk}. 

The elliptic flow has also been measured at RHIC 
for different hadron species, such as pions, kaons, nucleons and $\Lambda$ 
\cite{Adler:2001nb,Adler:2002pb}. 
The experimental data show that at low $\pt$ the elliptic flow of heavier 
particles is smaller than that of lighter particles. In the hydrodynamical 
model, this mass ordering of elliptic flow at low $\pt$ is attributed to the 
mass dependence of radial flow \cite{Huovinen:2001cy}. For high $\pt$ hadrons, 
we expect the flavor dependence to be different from that at 
low $\pt$, since high transverse momentum hadrons originate from 
hard processes while low transverse momentum particles are mostly produced
from soft non-perturbative processes and are much closer to thermal 
equilibrium. Indeed, the observed saturation of hadron elliptic flow
at $\pt>2$ GeV$/c$ \cite{Adler:2002pb,Adler:2002ct} contradicts the 
predictions from the hydrodynamical model \cite{Kolb:2000fh}, 
but is roughly consistent with the results expected from a large parton 
transport opacity \cite{Molnar:v2} or energy loss \cite{Gyulassy:2001gk} 
in the partonic matter. 

In this Letter, we shall study the flavor dependence of the elliptic flow 
of high $\pt$ hadrons in ultra-relativistic heavy ion collisions,
using a quark coalescence model to describe the phase transition 
from the partonic matter to the hadronic matter. In this model, 
the elliptic flow of high $\pt$ hadrons can be expressed in terms of 
the elliptic flow of high $\pt$ quarks. As a result, several relations 
between the elliptic flow of hadrons of different flavors are obtained.
We further discuss some special cases where these relations become more 
transparent. Throughout this study, we limit the discussions to 
hadrons made of $u$, $d$, $s$ quarks and antiquarks.

In the quark coalescence model, one assumes that quarks and antiquarks 
are the effective degrees of freedom in the parton phase near the phase 
transition, and they combine to form hadrons according to the 
valence quark structure of hadrons. A meson is thus formed from the 
coalescence of a quark and an antiquark, while a baryon is due to 
the coalescence of three quarks. The idea of quark coalescence has been 
used in models such as the ALCOR \cite{Biro:1994mp} or MICOR model 
\cite{Csizmadia:1998vp} to describe hadron abundance and the AMPT model 
with string melting \cite{Lin:2001zk} to describe the elliptic flow at RHIC. 

In ultra-relativistic heavy ion collisions, high $\pt$ partons are produced
from initial hard scatterings between nucleons for which the perturbative QCD 
is applicable. From the leading-order calculation, the parton transverse 
momentum spectrum from the subprocess of a two-parton hard scattering 
is given by:
\begin{eqnarray}
\frac {d\sigma}{d\hat t} \propto \frac{1}{\pt^4}.
\end{eqnarray}
The high $\pt$ parton spectrum thus follows an inverse power law modulo the 
corrections from the parton distribution function in the nucleus 
and higher-order effects. On the other hand, low $\pt$ partons, that 
are produced from initial soft processes and dominate the dynamics of 
partonic evolution in heavy ion collisions at RHIC, typically have an 
exponential spectrum close to a thermal distribution. The parton 
$\pt$ spectrum can thus be represented by an exponential function below 
a certain momentum scale $p_0$ and an inverse power law above $p_0$. 

Let us consider via the quark coalescence model the formation of a high 
$\pt$ meson with transverse momentum $\vec p_H$ from one parton with 
$\vec p_H$ and one parton with zero $\pt$, or from two partons with equal 
high $\pt$ of $\vec p_H/2$. The ratio of the probabilities for forming a 
high $\pt$ meson in these two cases is then proportional to 
$ \left ( e \pt /4 p_0 \right)^n$, where $n$ represents the exponent 
of the inverse power law for final high $\pt$ partons. 
Since this ratio is much greater than one for $\pt\gg p_0$, 
a high $\pt$ meson is dominantly formed from the coalescence of one 
high $\pt$ parton and one soft parton. Similarly, a high $\pt$ baryon 
is mainly formed from the coalescence of one high $\pt$ parton and 
two soft partons. 

The transverse momentum distribution $F(\vpp)=dN/(dp_x dp_y)$ 
of initial high $\pt$ mesons formed after the phase transition can thus 
be expressed in terms of that of final high $\pt$ partons as 
\begin{eqnarray}
F_H(\vpp)=F_i(\vpp) c_j + F_j(\vpp) c_i, 
\label{fh}
\end{eqnarray}
where $i$ and $j$ denote the flavor of the valence quark and antiquark 
of meson $H$.  The coefficient $c_i$ represents the capture probability 
for a soft parton $i$ by a high $\pt$ parton to form a high $\pt$ meson;  
it is thus related to the density of soft quarks near the phase transition. 
For high $\pt$ baryons or antibaryons, one can write down a 
similar expression, involving the product of two $c_i$'s,
for their transverse momentum distribution. 

The elliptic flow is generated during the early stage of heavy ion 
collisions when the pressure gradient and the spatial azimuthal asymmetry 
are the largest \cite{Sorge:1996pc,Teaney:v2,Zhang:1999rs,Lin:2001zk}. 
In transport model studies, it has been found that the elliptic flow 
in heavy ion collisions at RHIC develops mostly in the initial partonic 
phase, with later hadronic interactions having negligible effects on its 
final value \cite{Teaney:v2,Lin:2001zk}. We expect that the elliptic flow 
of high $\pt$ hadrons are even less affected by hadronic interactions, 
as the proper formation time from a high $\pt$ parton to a hadron 
is increased by a large Lorentz boost factor in the laboratory frame, 
leading to a much lower hadronic density when high $\pt$ hadrons 
are formed. We can thus use Eq.~(\ref{fh}) to relate the {\em final} 
elliptic flow of high $\pt$ hadrons to that of high $\pt$ partons \cite{mt}.  
For mesons, we have
\begin{eqnarray}
v_{2,H}(\pt)&=&\frac{\int \cos (2\phi^\prime) F_H(\vpp) 
d\phi^\prime} {\int F_H(\vpp) d\phi^\prime} 
\nonumber \\
&=&\frac{v_{2,i}(\pt) f_i(\pt) c_j + v_{2,j}(\pt) f_j(\pt) c_i}
{f_i(\pt) c_j + f_j(\pt) c_i},  
\end{eqnarray}
where $\phi^\prime$ is the azimuthal angle with respect to the 
reaction plane, and $f(\pt)=dN/(2\pi \pt d\pt)$ denotes the 
transverse momentum distribution after averaging over the azimuthal angle. 
In the following, we omit the label $\pt$ in the variables  
$v_2(\pt)$ and $f(\pt)$ but keep in mind that they are evaluated at 
a given high $\pt$. 

For SU(3) hadrons consisting of $u$, $d$, $s$ quarks and antiquarks, 
their $v_2$ values at high $\pt$ are then given~by:
\begin{eqnarray}
&&v_{2,\pi^+}\!\!=\!\!\frac{v_{2,u} f_u c_{\db} + v_{2,\db} f_{\db} c_u} 
{f_u c_{\db} + f_{\db} c_u},
~~v_{2,K^+}\!\!=\!\!\frac{v_{2,u} f_u c_{\sbar} + v_{2,\sbar} f_{\sbar} c_u}
{f_u c_{\sbar} + f_{\sbar} c_u},\nonumber \\
&&v_{2,\phi}=\frac{v_{2,s} f_s c_{\sbar} + v_{2,\sbar} f_{\sbar} c_s}
{f_s c_{\sbar} + f_{\sbar} c_s}, \nonumber \\
&&v_{2,p}=\frac{v_{2,u} f_u c_d + v_{2,d} f_d c_u/2}
{f_u c_d + f_d c_u/2}, \nonumber \\
&&v_{2,\Lambda}=v_{2,\Sigma^0}=\frac{v_{2,u} f_u c_d c_s + v_{2,d} f_d c_u c_s
+ v_{2,s} f_s c_u c_d}
{f_u c_d c_s + f_d c_u c_s + f_s c_u c_d},\nonumber  \\
&&v_{2,\Xi^0}=\frac{v_{2,u} f_u c_s/2 + v_{2,s} f_s c_u}
{f_u c_s/2 + f_s c_u}, ~~v_{2,\Omega}=v_{2,s},
\label{v2h}
\end{eqnarray}
with similar expressions for isospin partners and antiparticles. 

The above relations become simpler if the quantities
$v_{2,i}, f_i$, and $c_i$ are independent of isospin and are also the same
for strange and antistrange quarks,
i.e., $u=d \equiv q,~\ub=\db \equiv \qb$, and $s=\sbar$. 
These conditions are approximately satisfied in heavy ion collisions at RHIC 
as the $\pi^+/\pi^-$ ratio is almost one around central rapidity 
\cite{Adcox:2001mf,Phobos,valence}. 
In this isospin symmetric and strange-antistrange symmetric limit, 
the $v_2$ values for hadrons at a given high $\pt$ are given by:
\begin{eqnarray}
&&v_{2,N}=v_{2,q},~~v_{2,\bar N}=v_{2,\qb},~~ 
v_{2,\phi}=v_{2,\Omega}=v_{2,s}, \nonumber \\
&&v_{2,\pi^+}=v_{2,\pi^0}=v_{2,\pi^-}=
\frac{v_{2,q} + r_{\qb} v_{2,\qb}}{1+r_{\qb}}, 
\nonumber \\
&&v_{2,K^+}=\frac{v_{2,q} + r_s v_{2,s}}{1+r_s}, ~~
v_{2,K^-}=\frac{v_{2,\qb} + r_s v_{2,s}/r_{\qb}}{1+r_s/r_{\qb}}, \nonumber \\
&&v_{2,\Lambda}=v_{2,\Sigma}=\frac{2v_{2,q} + r_s v_{2,s}}{2+r_s}, \nonumber \\
&&v_{2,\bar \Lambda}=v_{2,\bar \Sigma}=\frac{2v_{2,\qb} 
+ r_s v_{2,s}/r_{\qb}}{2+r_s/r_{\qb}}, \nonumber \\
&&v_{2,\Xi}=\frac{v_{2,q}/2 + v_{2,s}r_s}{1/2+r_s}, 
~~v_{2,\bar \Xi}=\frac{v_{2,\qb}/2 + v_{2,s}r_s/r_{\qb}}{1/2+r_s/r_{\qb}}, 
\label{issym}
\end{eqnarray}
with $N$ denoting a nucleon. In the above, the $\pt$-dependent variables 
$r_{\qb}$ and $r_s$ are defined as
\begin{eqnarray}
r_{\qb}=\frac {f_{\qb} c_q}{f_q c_{\qb}},~~r_s=\frac {f_s c_q}{f_q c_s}.
\label{rdef}
\end{eqnarray} 
From Eq.(\ref{issym}), we see that, e.g., $r_{\qb}$ can be determined from 
the $v_2$ of high $\pt$ pion, proton and antiproton, while $r_s$ can be
determined from the $v_2$ of high $\pt$ proton, kaon, and $\phi$ meson. 

Since the $K^+/K^-$ ratio is close to one and $\bar p/p$ ratio is 
about 0.7 in heavy ion collisions at RHIC \cite{Adcox:2001mf,Phobos,valence}, 
and they should be closer to one in heavy ion collisions at LHC, 
we consider the case where the variables $v_{2,i}, f_i$ and $c_i$ are 
the same for quarks and antiquarks, i.e., $q=\qb$ and thus $r_{\qb}=1$. 
For such a quark-antiquark symmetric partonic matter, 
Eq.(\ref{issym}) simplifies to:
\begin{eqnarray}
&&v_{2,\pi}=v_{2,N}=v_{2,q},~~
v_{2,\phi}=v_{2,\Omega}=v_{2,s}, 
\label{v2sym1} \\
&&v_{2,K}=\frac{v_{2,q} + r_s v_{2,s}}{1+r_s}, ~~
v_{2,\Lambda}=v_{2,\Sigma}=\frac{2 v_{2,q} + r_s v_{2,s}}{2+r_s}, 
\nonumber \\
&&v_{2,\Xi}=\frac{v_{2,q} + 2r_s v_{2,s}}{1+2r_s}.
\label{v2sym2}
\end{eqnarray}

Eliminating the variable $r_s$ in Eq.~(\ref{v2sym2}), 
we obtain two relations involving the $v_2$ values of four 
different hadron species, and they can be any two of the following 
three relations:
\begin{eqnarray}
&&\left (v_{2,\pi}\!\!-\!\!v_{2,K} \right )
\cdot \left (v_{2,\Lambda}\!\!-\!\!v_{2,\phi} \right ) 
=2 \left (v_{2,\pi}\!\!-\!\!v_{2,\Lambda} \right )
\cdot \left (v_{2,K}\!\!-\!\!v_{2,\phi} \right ), \nonumber \\
&&2 \left (v_{2,\pi}\!\!-\!\!v_{2,K} \right )
\cdot \left (v_{2,\Xi\!\!}-\!\!v_{2,\phi} \right )
= \left (v_{2,\pi}\!\!-\!\!v_{2,\Xi} \right )
\cdot \left (v_{2,K}\!\!-\!\!v_{2,\phi} \right ), \nonumber \\
&&\left (v_{2,\pi}\!\!-\!\!v_{2,\Xi} \right )
\cdot \left (v_{2,\Lambda}\!\!-\!\!v_{2,\phi} \right )
= 4 \left (v_{2,\pi}\!\!-\!\!v_{2,\Lambda} \right )
\cdot \left (v_{2,\Xi}\!\!-\!\!v_{2,\phi} \right ).
\label{four}
\end{eqnarray}

These relations on the elliptic flow of hadrons of different flavors
become even simpler in several limits for the value of $r_s$. 
In the limit of $r_s \rightarrow 0$ due to $f_s/f_q \rightarrow 0$, i.e., 
if there are very few high $\pt$ strange quarks relative to light quarks, 
we have 
$v_{2,\pi} = v_{2,K} = v_{2,N}
=v_{2,\Lambda}=v_{2,\Sigma}=v_{2,\Xi}=v_{2,q},~
v_{2,\phi}=v_{2,\Omega}=v_{2,s}$.
This is simply due to the fact that all strange hadrons with
light valence quarks consist of leading light quarks.
In the opposite limit of $r_s \rightarrow \infty$, i.e., if 
the number of high $\pt$ strange quarks is much larger than that of 
light quarks or the capture probability of a soft strange quark is 
much smaller than that of a soft light quark, all strange hadrons 
with light valence quarks consist of leading strange quarks. 
In this case, we have
$v_{2,\pi}=v_{2,N} =v_{2,q}, ~
v_{2,K}=v_{2,\phi}= v_{2,\Lambda}=v_{2,\Sigma}=v_{2,\Xi}
=v_{2,\Omega}=v_{2,s}$.

Another interesting limit is $r_s=1$, which would be the case if 
the spectrum of high $\pt$ strange quarks is the same as that of light quarks, 
and the capture probability of a soft strange quark is the same 
as that of a soft light quark, or even though the above two factors are 
different but they cancel each other. In this limit, Eq.(\ref{v2sym2}) gives 
\begin{eqnarray}
v_{2,K}\!\!=\!\!\frac{v_{2,q} \!\!+\!\! v_{2,s}}{2},~
v_{2,\Lambda}\!\!=\!\!\frac{2v_{2,q}\!\! +\!\! v_{2,s}}{3},~
v_{2,\Xi}\!\!=\!\!\frac{v_{2,q} \!\!+\!\! 2 v_{2,s}}{3}.  
\label{rs1}
\end{eqnarray}
These relations, together with Eq.(\ref{v2sym1}), 
show that the $v_2$ values of hadrons at high $\pt$ follow a 
simple quark flavor counting rule when $r_s=1$. 

Eqs.~(\ref{v2sym1}-\ref{v2sym2}) show that 
the dependence of the elliptic flows of high $\pt$ hadrons
on their flavor composition is determined by the relative magnitude of the
elliptic flow of high $\pt$ strange quarks to that of high $\pt$
light quarks.  If strange quarks have the same elliptic flow as 
light quarks at high $\pt$, i.e., $v_{2,s}=v_{2,q}$, then $v_{2,H}=v_{2,q}$ 
for all SU(3) hadrons regardless of the value of $r_s$. 
This is true even if the $\pt$ spectrum for strange quarks is different 
from that for light quarks. We note that in the present study 
we are only concerned with the {\em relative} magnitude of the 
elliptic flow of different hadrons at high $\pt$, not their absolute 
magnitudes or shape.

On the other hand, fast moving heavy quarks have been shown to 
suffer less energy loss in a thermalized parton plasma than fast moving
light quarks \cite{Braaten:1991we,Thoma:1990fm,Dokshitzer:2001zm}. 
In the parton transport model, this would imply that high $\pt$ strange 
and heavier quarks may have smaller scattering cross sections 
than light quarks in a partonic matter. 
It is thus possible that the elliptic flow of strange 
quarks is smaller than that of light quarks, i.e., $v_{2,s}<v_{2,q}$. 
In this case, we obtain from Eqs.~(\ref{v2sym1}-\ref{v2sym2}) the 
following flavor ordering of the $v_2$ values for hadrons at a given
high~$\pt$:
\begin{eqnarray}
(v_{2,\pi}\!\! =\!\! v_{2,N} ) 
\!>\! (v_{2,\Lambda}\!\!=\!\!v_{2,\Sigma} ) 
\!>\! v_{2,K} \!>\! v_{2,\Xi} \!>\! (v_{2,\phi}\!\!=\!\!v_{2,\Omega} ).
\label{order}
\end{eqnarray}

\begin{figure}[h]
\centerline{\epsfig{file=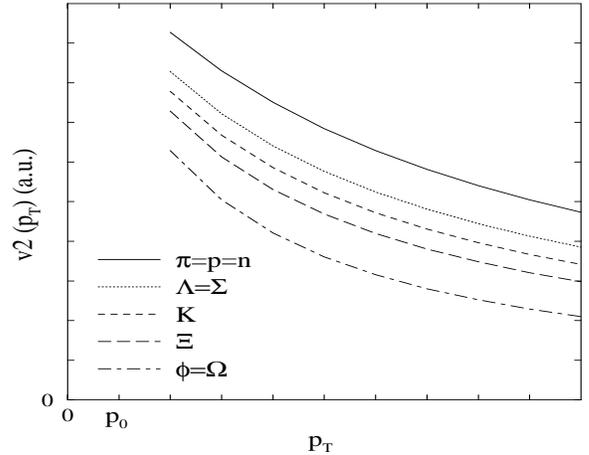,width=3in,height=2.4in,angle=0}}
\vspace{0.3cm}
\caption{Schematic plot for the flavor ordering of the elliptic
flows of hadrons at high $\pt$. Details are given in the text.}
\label{figv2}
\end{figure}

In Fig.~\ref{figv2}, we illustrate the flavor ordering of hadron
elliptic flows at high $\pt$ for the case of $v_{2,s}<v_{2,q}$. 
The spacings between different curves 
correspond to the case of $r_s=1$ and thus follow the quark counting 
relations of Eqs.(\ref{v2sym1}) and (\ref{rs1}). 
We note that the vertical scale for $v_2$ is in arbitrary units, 
and the shape of $v_2$ as a function of $\pt$ is also arbitrary. 
The scale $p_0$ denotes the typical transverse momentum 
above which the $\pt$ spectra of final partons changes from soft to hard,  
and its value should probably be a few GeV$/c$. 
All curves are shown well above $p_0$, reflecting the fact that the 
relations derived in the present study only apply to hadron elliptic 
flows at high $\pt$.  In general, $r_s$ can take any finite positive value,
but the flavor ordering of hadron elliptic flows remains similar to 
that shown in Fig.~\ref{figv2} as long as $v_{2,s}<v_{2,q}$. 
However, the spacings between different curves can be different, while 
still being constraint by the two relations given in Eq.(\ref{four}).
Since the $v_2$ magnitudes of hadrons follow the mass ordering at low $\pt$ 
and the flavor ordering at high $\pt$, the curve for kaon $v_2(\pt)$,
which is above those for proton and $\Lambda$ at low $\pt$, 
will cross and become lower than the latter two curves as $\pt$ increases.
A similar relation exists between the curve for $\phi$ meson $v_2(\pt)$
and those for $\Lambda$ and $\Xi$. 

We have not included the effects of resonance decays in this 
study. The relations shown in Eq.(\ref{v2h}) can be extended 
to resonances such as $\eta$, $\rho$, $\omega$, $K^*$, and $\Delta$. 
These resonances at high $\pt$ will decay to stable hadrons at different
transverse momenta, thus complicating the relations we have thus derived
for hadrons which are directly formed from the quark coalescence. 
Since the transverse momentum of a decay product is usually small 
compared to that of the parent hadron, and the inverse power law 
spectrum shows a rapid decrease with $\pt$, we expect that the 
resonance contribution to hadron elliptic flow at high $\pt$ is small 
compared to the contribution from directly formed hadrons. 

In summary, using a parton coalescence model to describe the 
formation of hadrons from the initial partonic matter in ultra-relativistic
heavy ion collisions, we have studied the dependence of the 
elliptic flows of hadrons at high $\pt$ on their flavor composition. 
Since the elliptic flow is generated mostly in the early partonic phase, 
and high $\pt$ hadrons are mainly formed from the coalescence of a high 
$\pt$ quark or antiquark produced from the initial hard processes and 
low $\pt$ quarks or antiquarks from the soft processes, 
the magnitudes of the hadron elliptic flows at high $\pt$ 
are determined by that of high $\pt$ quarks.
The relations between hadron and parton elliptic flows at high $\pt$ 
also depend on the final quark spectrum at high $\pt$ ($f_i(\pt)$)  
and the capture probability of a soft quark ($c_i$) by a high $\pt$ quark 
to form a high $\pt$ hadron. If strange quarks have a smaller elliptic 
flow than the light quarks, then the quark coalescence model 
leads to the flavor ordering in the elliptic flows of the hadrons 
formed from an isospin symmetric and quark-antiquark symmetric partonic matter,
i.e.,  
$(v_{2,\pi}\!\! =\!\! v_{2,N} ) 
> (v_{2,\Lambda}\!\!=\!\!v_{2,\Sigma} ) 
> v_{2,K} > v_{2,\Xi} > (v_{2,\phi}\!\!=\!\!v_{2,\Omega})$.
We have also obtained two relations which are independent 
of $f_i(\pt)$ and $c_i$ and involve the elliptic flows of four 
hadron species at high $\pt$.  In the special case that $f_i(\pt)$ 
and $c_i$ are the same for strange quarks and light quarks, values of
the elliptic flows for high $\pt$ hadrons of different flavors are 
found to follow the quark counting rule. It will be very interesting 
to test these predictions in current and future heavy ion collisions. 
Such studies will provide valuable information on whether a partonic 
matter is formed in the collisions and the subsequent formation of 
hadrons can be described by the quark coalescence model.

\medskip

We appreciate useful discussions with H. Huang, P. Sorensen, and A. Tai. 
This paper is based on work supported by the U.S. National Science 
Foundation under Grant Nos. PHY-9870038 and PHY-0098805, the Welch 
Foundation under Grant No. A-1358, and the Texas Advanced Research 
Program under Grant No. FY99-010366-0081.


\end{multicols}


\begin{thebibliography}{99}
\bibitem{mt}
We caution that Eq.~(\ref{fh}) cannot be used for 
observables such as the final hadron $m_{\rm T}$ spectra, as
they are significantly modified by interactions in the 
hadronic phase.
\bibitem{valence}
We note that both the isospin symmetry and quark-antiquark symmetry 
will break down for quarks with $\pt$ above the scale of the order of 
$\sqrt s/6$, with $\sqrt s$ being the center-of-mass energy, 
since the valence quark distribution function of nucleons starts 
to dominate the flavor of the produced jets.
\end{thebibliography}
\end{document}